\documentstyle[12pt]{article}
\begin{document}
\begin{center}
{\large \bf HIGH-ENERGY NUCLEAR PHYSICS} \\[0pt] \vspace{0.1cm} 
{\large \bf WITH LORENTZ SYMMETRY VIOLATION}\\[0pt]
\vspace{0.5cm} {\bf Luis GONZALEZ-MESTRES}\footnote{%
E-mail: lgonzalz@vxcern.cern.ch}
\\[0pt]
\vspace{0.3cm}
{\it Laboratoire de Physique Corpusculaire, Coll\`ege de France \\
11 pl.
Marcellin-Berthelot, 75231 Paris Cedex 05, France
\\[0pt] and
\\[0pt]
Laboratoire d'Annecy-le-Vieux de Physique des Particules \\ B.P. 110 , 74941 
Annecy-le-Vieux Cedex, 
France} \vspace{0.8cm}
\end{center}

\begin{abstract}
If textbook Lorentz invariance is actually
a property of the equations describing a sector
of the excitations of vacuum above some critical distance scale,
several sectors of matter with different
critical speeds in vacuum can coexist and an absolute rest frame (the vacuum
rest frame)
may exist without contradicting the apparent Lorentz invariance felt by
"ordinary" particles (particles with critical speed in vacuum equal to $c$ ,
the speed of light). Sectorial Lorentz invariance, reflected by the fact that
all particles of a given dynamical sector have the same critical speed in
vacuum, will then be an expression of a fundamental sectorial symmetry
(e.g. preonic grand unification or extended supersymmetry) protecting a
parameter of the equations of motion. Furthermore, the sectorial Lorentz
symmetry may be only a low-energy limit, in the same way as the relation
$\omega $ (frequency) = $c_s$ (speed of sound) $k$ (wave vector) holds for
low-energy phonons in a crystal. In this context, phenomena
such as the absence of Greisen-Zatsepin-Kuzmin cutoff for protons and nuclei
and the stability
of unstable particles (e.g. neutron, several nuclei...)
at very high energy are basic properties of a wide
class of noncausal models where local Lorentz invariance is broken
introducing a fundamental length. Observable phenomena are expected
at very short wavelength scales, even if
Lorentz symmetry violation remains invisible to standard low-energy
tests. We present a detailed discussion of the implications of Lorentz symmetry
violation for very high-energy nuclear physics.
\end{abstract}

\section{Introduction}

In previous papers (Gonzalez-Mestres, 1997a and 1997b) we suggested
that, as a consequence of nonlocal dynamics at Planck scale or at some other
fundamental length scale, Lorentz symmetry violation can result in a
modification of the equation relating energy and momentum which
would write in the vacuum rest frame:
\equation
E~~=~~(2\pi )^{-1}~h~c~a^{-1}~e~(k~a)
\endequation
where $E$ is the energy of the particle,
$h$ the Planck constant, $c$ the speed of light,
$a$ a fundamental length scale (that we can
naturally identify with the Planck length, but other choices of the
fundamental distance scale are possible), $k$
the wave vector modulus and
$[e~(k~a)]^2$ is a convex
function of $(k~a)^2$ obtained from nonlocal vacuum dynamics.

Rather generally, we find
that, at wave vector scales below the inverse of the fundamental length scale,
Lorentz symmetry violation in relativistic kinematics can be parameterized
writing:
\equation
e~(k~a)~~\simeq ~~[(k~a)^2~-~\alpha ~(k~a)^4~+~(2\pi ~a)^2~h^{-2}~m^2~c^2]^{1/2}
\endequation
where $\alpha $ is a positive
constant between $10^{-1}$ and $10^{-2}$ if the nonlocal effect 
occurs at leading level at the fundamental length scale
(which we hereafter assume unless the contrary is explicitly stated). 
At high energy, we can write:
\equation
e~(k~a)~~\simeq ~~k~a~[1~-~\alpha ~(k~a)^2/2]~
+~2~\pi ^2~h^{-2}~k^{-1}~a~m^2~c^2
\endequation
and, in any case, we expect observable kinematical effects when the 
deformation term
$\alpha (ka)^3/2$ becomes as large as the mass term
$2~\pi ^2~h^{-2}~k^{-1}~a~m^2~c^2$ .
Assuming that, apart form the value of the mass, expression (2) is universal
for all existing particles whose critical speed in vacuum is equal to the speed 
of light in the Lorentz-invariant limit
(i.e. assuming universal values of $c$ and $\alpha$) , 
we found several important effects
(Gonzalez-Mestres, 1997c) that can turn the study of very high-energy cosmic 
rays into a unique frame to test special relativity and physics at Planck
scale:

a) The Greisen-Zatsepin-Kuzmin (GZK) cutoff on very high-energy cosmic
protons and nuclei (Greisen, 1966;
Zatsepin and Kuzmin, 1966) does no longer apply (Gonzalez-Mestres,
1997a and 1997b). Very high-energy cosmic rays originating
from most of the presently
observable Universe can reach the earth and generate the highest-energy
detected events. Indeed, fits to data below $E~=~10^{20}~eV$ using standard
relativistic kinematics (e.g. Dova, Epele and Hojvat, 1997) predict a sharp
fall of the event rate at this energy, in contradiction with data
(Bird et al., 1993 and 1996; Hayashida et al., 1994
and 1997; Yoshida et al., 1995)
which suggest that events above $10^{20}~eV$ are produced at a
significant rate. Lorentz symmetry violation from physics at Planck scale
provides a natural way out.

b) Unstable particles with at least two massive particles in the final state
of all their decay channels 
(neutron, nuclei, $\Delta ^{++}$ , possibly muons and $\tau $'s...)
become stable at very high energy
(Gonzalez-Mestres, 1997a and 1997b).
In any case, unstable particles live longer than naively expected with exact
Lorentz invariance and, at high enough energy,
the effect becomes much stronger than previously estimated for nonlocal models
(Anchordoqui, Dova, G\'omez Dumm
and Lacentre, 1997)
ignoring the small violation of relativistic kinematics. Not only particles 
previously discarded because of their lifetimes can be candidates for the
highest-energy cosmic-ray events, but very high-energy
cascade development can be modified
(for instance, if the $\pi ^0$ lives longer at energies above $\approx ~
10^{18}~eV$, thus favoring hadronic interactions and muon pairs 
and producing less 
electromagnetic showers).

c) The allowed final-state phase space of two-body collisions is seriously
modified at very high energy, especially when, in the vacuum rest frame
where expressions (1) - (3) apply,
a very high-energy particle
collides with a low-energy target (Gonzalez-Mestres, 1997d). Energy
conservation reduces the final-state
phase space at very high energy and can lead to a sharp fall of cross sections
starting at incoming-particle wave vectors well below the
inverse of the fundamental length, essentially above 
$E~\approx ~(E_T~a^{-2}~h^2~c^2)^{1/3}$ where $E_T$ is the energy of the 
low-energy target. For $a~\approx ~10^{-33}~cm$ , this scale corresponds to:
$\approx ~10^{22}~eV$ if the target is a rest proton; $\approx ~10^{21}~eV$ if 
it is a rest electron; $\approx ~10^{20}~eV$ for a $\approx ~1~keV$ photon,
and $\approx ~10^{19}~eV$ if the target is a visible photon.  
For a proton impinging on a $\approx ~10^{-3}~eV$ photon from cosmic microwave
background radiation, and taking $\alpha ~\approx ~1/12$ as in models based on
an isotropic extension of the Bravais lattice dynamics (Gonzalez-Mestres, 
1997a), we expect the fall of cross sections to occur above $E~\approx ~
5.10^{18}~eV$, the critical energy where the derivatives of the mass term 
$m^2~c^3~p^{-1}/2$ and of the deformation term $\alpha ~p~(k~a)^2/2$  
become equal in the expression relating the proton
energy $E$ to its momentum $p$ . With $a~\approx ~10^{-30}~cm$ , still allowed
by cosmic-ray data (Gonzalez-Mestres, 1997d), the critical energy scale 
can be as low as
$E~\approx ~10^{17}~eV$ ; with $a~\approx ~10^{-35}~cm$ (still compatible with
data in the region of the predicted GZK cutoff), it would be at
$E~\approx ~5.10^{19}~eV$ .
Similar considerations lead to a fall of
radiation under external forces (e.g. synchrotron radiation) above this
energy scale. In the case of
a very high-energy $\gamma $ ray, taking $a~\approx ~10^{-33}~cm$ , 
the deformed relativistic
kinematics inhibits collisions with  $\approx ~10^{-3}~eV$ photons
from cosmic background radiation above $E~\approx ~10^{18}~eV$~,
with $\approx ~10^{-6}~eV$ photons above 
$E~\approx ~10^{17}~eV$ and with $\approx ~10^{-9}~eV$ photons above 
$E~\approx ~10^{16}~eV$~. Taking $a~\approx ~10^{-30}~cm$ would lower these
critical energies by a factor 100 
according to the previous formulae, 
whereas the choice $a~\approx 10^{-35}~cm$ would raise
them by a factor of 20 . Obviously, any relevant 
high-energy astrophysical
calculation must consider the possibility that Lorentz symmetry be violated at
some fundamental length scale. 
  
d) In astrophysical processes,
the new kinematics may inhibit phenomena such as GZK-like cutoffs,
photodisintegration of nuclei, decays,
radiation emission under external forces
(similar to a collision with a very low-energy target), momentum loss
(which at very high energy does not imply deceleration) through collisions,
production of lower-energy secondaries... {\bf potentially solving all
the basic
problems raised by the highest-energy cosmic rays}
(Gonzalez-Mestres, 1997c and 1997d). Due to the fall of
cross sections, energy losses become much weaker than expected with
relativistic kinematics and astrophysical particles can be pushed to much
higher energies (once energies above $10^{17}~eV$ have been reached through
conventional mechanisms, synchrotron radiation and collisions with ambient
radiation may start to be
inhibited by the new kinematics); similarly, astrophysical
particles will be able to propagate to
much longer astrophysical distances, and many more sources
(in practically all the presently observable Universe) can 
produce very high-energy cosmic rays reaching the
earth; as particle lifetimes are
much longer, new possibilities arise for the nature of these cosmic rays.
The same considerations apply to nuclei, but their composite nature requires
a specific discussion. Models of very high-energy
astrophysical
processes cannot ignore a possible Lorentz symmetry violation
at Planck scale, in which case observable effects are predicted for the
highest-energy detected particles.

e) If the new kinematics can explain the existence of $\approx 10^{20}~eV$
events, it also predicts that, above some higher energy
scale (around $\approx 10^{22}~eV$ for $a~\approx ~10^{-33}~cm$), the fall of
cross sections will prevent the cosmic ray from depositing most of its
energy in the
atmosphere (Gonzalez-Mestres, 1997d).
Such extremely high-energy particles will produce atypical events of
apparently much lower energy. New analysis
of data and experimental designs are required to explore
this possibility.

Velocity reaches its maximum at 
$k~\approx ~(8\pi ^2~\alpha ^{-1})^{1/4}~(m~c~h^{-1}~a^{-1})^{1/2}$ .
Above this value, increase of momentum amounts to deceleration.
In our ansatz, 
observable effects of local Lorentz invariance breaking
arise, at leading level, well below the critical
wavelength scale $a^{-1}$ due to the fact that, contrary to previous models 
(f.i. R\'edei, 1967), we directly apply non-locality to particle 
propagators and not only 
to the interaction hamiltonian. In contrast with 
previous
patterns (f.i. Blokhintsev, 1966), $s-t-u$ kinematics ceases to make sense
and the motion of the global system with respect to the vacuum rest frame plays
a crucial role. The physics of elastic two-body scattering will depend  
on five kinematical variables.
Noncausal dispersion relations (Blokhintsev and Kolerov, 
1964) should be reconsidered, taking into account the departure from 
relativistic kinematics. As previously stressed (Gonzalez-Mestres,
1997a) , this apparent
nonlocality may actually reflect the existence of superluminal sectors of
matter (Gonzalez-Mestres, 1996) where causality would hold at the
superluminal level (Gonzalez-Mestres, 1997e). Indeed, electromagnetism 
appears as a nonlocal interaction in the Bravais model of phonon dynamics,
due to the fact that electromagnetic signals propagate much faster than
lattice vibrations.

In this note, we would like to discuss the consequences of the
new kinematics for very high-energy nuclear physics, focusing on the 
application of the deformed dispersion relation to situations where the
particle (proton, nucleus...) is actually a composite object. As compared 
to a previous attempt to extend to composite and macroscopic objects
a different version of deformed special relativity (Bacry, 1993), our paper
focuses more on the understanding of the physical content of the kinematics 
instead of trying to
directly extend to all bodies the same underlying deformed Poincar\'e algebra.


\section{Deformed kinematics and composite objects}

A basic question is how deformed relativistic kinematics applies to 
composite objects. By "elementary" particles we 
should in principle mean those that are
generated by fundamental dynamics at Planck scale or beyond this scale
and which, at this level, can be described similar to phonons in a
lattice (Gonzalez-Mestres, 1997a) or to solitons or
topological defects in a medium.
This can be the case of quarks, leptons and gauge bosons or, instead,
of same preons which would be the constituents of these particles.
Because of quark confinement, it can also be the case of hadrons.
Supersymmetric extensions of these criteria are obvious. A natural
hypothesis, that we have used in previous calculations, 
may be that $c$ and $\alpha $ have universal values for
all "elementary" particles and also for nuclei. But, as we shall see below, 
the situation for hadrons appears to be less trivial than 
suggested by this naive 
formulation and the hypothesis may turn out to be fundamentally wrong for
heavy nuclei. The transition from nucleons to nuclei will be a crucial test
for all models. 

Standard relativity extends trivially from microscopic to macroscopic
objects, from constituents to composite entities. When boosting any
physical body initially at rest, its rest energy is multiplied by the
Lorentz factor $\gamma ~=~(1~-~v^2c^{-2})^{-1/2}$ for a velocity
${\vec v}$ and the momentum is
given by the galilean momentum times $\gamma $ . This operation is 
additive with respect to the constituents, 
and works in the same way for elementary and composite
physical systems. Schematically, relativistic composite models admit a "parton" 
description where each constituent carries 
a fraction of the total energy 
and momentum, and this energy-independent fraction (neglecting 
radiative corrections) 
can vary
according to a probability distribution given by the basic dynamics.
Formally, we can split the rest energy $M~c^2$ of the composite system
in $N$ parts $M_i~c^2$ ($i~=~1,...,N$ , $M_1~+~...~+M_N~=M$) and attribute 
to each part: a) an energy $E_i~=~M_i~c^2~\gamma $ ; b) a momentum  
${\vec {p_i}}~=~M_i~{\vec v}~\gamma $~. The fraction of energy and momentum
carried by the $i$-th parton is $M_i/M$ .
Then, the speed $v~=~dE/dp$ is 
independent of the fraction of the total momentum and energy carried by
the parton: all partons can have the same velocity, equal to that of the
composite model, 
independently of how they share the total energy and momentum.
This property, essential to the stability of 
the composite system, is due to the absence of any distance
or mass scale other than the total mass (the formal "rest energies" of
the partons being fractions of the total rest energy).
The deformation of special relativity 
modifies this situation at very high energy
by the introduction of the fundamental scale $a$ associated to nonlocal 
interactions, and produces leading effects well below 
$E~\approx ~(2\pi ~a)^{-1}~h~c$ . 

At very high energy, the speed of an "elementary" particle 
in deformed relativity is given
by (Gonzalez-Mestres, 1997a):
\equation
v~~=~~c~~[de~(k~a)~/~d(k~a)]
\endequation
which, for $e~(k~a)~=~[sin^2~(k~a)~+~(2\pi ~a)^2~h^{-2}~m^2~c^2]^{1/2}$ 
and $m~\neq ~0$ ,
falls to zero as $k$ approaches the value $\pm \pi /2a$ .
The velocity vector changes sign at these values of $k$ and tends again 
to zero as $k$ approaches $\pm \pi /a$ . 
Similar properties, with slightly different formulae, 
are found in most nonlocal models.
A basic property of deformed 
relativistic kinematics is that: a) the velocity of each 
parton depends on the fractions it carries of the total energy and momentum; 
b) the
fractions of the total energy and momentum carried by a given parton are not 
exactly
identical; c) for a given value 
of the total momentum, the total energy depends on the way momentum is shared 
by the partons. At $TeV$ energies, and taking $a$ close to Planck scale, 
these effects are very small and can be
compensated by very small deviations from the standard parton picture.
However, at the highest energies reached by cosmic-ray events, the fluctuations
of the total energy due to the combination of standard structure 
functions with deformed relativistic kinematics for partons 
would become of the same
order as the mass term of the composite system and produce therefore a leading
effect. At such energies, we do not expect the parton model to remain
valid. With respect to standard relativity, a fundamental change has been
operated by the presence of an energy scale (the Planck scale, or any other
similar scale) other than the total rest energy and whose effect does not 
amount to a simple redefinition of partons. The new physics manifests
itself at wavelengths well below the inverse of the fundamental length scale, 
and can be experimentally checked at the highest observed cosmic-ray energies.

Assume that, inside a fast proton, quarks can be considered as "almost-free"
parton constituents to which deformed relativistic kinematics can be
individually applied.
At wave vectors above 
$k~\approx ~(8\pi ^2~\alpha ^{-1})^{1/4}~(m~c~h^{-1}~a^{-1})^{1/2}$ 
in the vacuum rest frame,
the velocity decreases as momentum increases and the term 
$\alpha ~(k~a)^3/2$ in (3) dominates over the term 
$(2\pi ~a)^2~h^{-2}~m^2~c^2$ ; 
no compensation is possible between the derivatives of the two terms.
At wave vectors well below 
$k~\approx ~(8\pi ^2~\alpha ^{-1})^{1/4}~(m~c~h^{-1}~a^{-1})^{1/2}$ ,
it would still be possible to compensate the small effect of the term 
$\alpha ~(k~a)^3/2$ in (3) by small corrections to the parton description: 
in order to preserve the collective propagation of the bound state, 
the fractions 
of the total energy and momentum carried by each parton would become
slightly velocity-dependent; in order to make the total energy independent
of the way momentum is shared by the partons, a momentum-dependent potential
term should be added. Above 
$k~\approx ~(8\pi ^2~\alpha ^{-1})^{1/4}~(m~c~h^{-1}~a^{-1})^{1/2}$ ,
the new kinematics produces a leading effect compelling all
partons to have almost the same energy and momentum.
Then, in a constituent model 
and due to the term $\alpha ~(k~a)^3/2$ which explicitly incorporates the
fundamental scale $a$ as the distance 
(or energy, or mass) scale, the constituents considered as
"almost-free" particles
would have the same velocity only if they carried almost exactly the same 
fractions of the total
momentum and energy. This definitely forces a departure from the parton model
and, since quarks cannot oscillate outside hadrons,
it suggests that in a stable hadron (where quarks are confined)
there exists only one motion (that of the overall system, instead of a set
of "almost-free" parton constituents)
to which the deformed quantum relativistic kinematics can be applied. 
Therefore, it indeed seems natural to extend to such systems the same 
formulae of deformed relativistic kinematics as 
for leptons and gauge bosons, with
the same values of $c$ and $\alpha $ . Simultaneously, we expect any parton 
model for hadrons to fail at energies above $\approx ~10^{19}~eV$ 
if $a~\approx 10^{-33}~cm$ , $\approx ~10^{20}~eV$ for 
$a~\approx 10^{-35}~cm$ 
and $\approx ~3.10^{17}~eV$ for 
$a~\approx 10^{-30}~cm$ . However, as will be seen below, soliton or
bound state dynamics can introduce deviations from the universality
of $\alpha $ .  

It must also be realized that, in a naive constituent 
picture, deformed relativistic
kinematics with universal values of $c$ and $\alpha $ cannot be applied
simultaneously to the constituents and to the composite system. If this
were the case, the former could by no means  
have the same velocity as the latter
at wave vectors above 
$k~\approx ~(8\pi ^2~\alpha ^{-1})^{1/4}~(m~c~h^{-1}~a^{-1})^{1/2}$ .
If deformed relativistic kinematics were to be applied simultaneously to
the constituents and to the composite system, the universality of 
$\alpha $ would have to be abandoned and a choice of $\alpha $ around
$\alpha _C~\approx ~N^{-2}~\alpha _P$ , where $\alpha _P$ is the value of
$\alpha $ for a single constituent and $N$ is the number of constituents,
would be required for the composite system. 
Nevertheless, such a choice seems to lack motivation for hadrons, as 
quarks are confined and in deformed relativistic kinematics the
internal structure of the 
composite system at very high energy does not allow for
several independent, large longitudinal momenta.

Similar questions can 
be heuristically considered from the point of view of Lorentz contraction
and dilation, using simplified soliton models. 
For instance, in the model based on 
an analogy with the one-dimensional 
Bravais lattice (Gonzalez-Mestres, 1997a), we can write the formal expansion
(see also Gonzalez-Mestres, 1997f):
\equation
\phi ~(x~\pm~a)~~-~~\phi ~(x)~~~=
~~~\Sigma _{n=0}^{\infty }~\phi ^{(n)}~(x)~~(\pm a)^n 
~(n!)^{-1}
\endequation
where $\phi ^{(n)}~=~d^n\phi /dx^n$ and 
$\phi ~(x)$ is the fixed-time 
wave function (extended from integer $x~a^{-1}$ to 
real $x$),
so that the expression $\phi ~(x)~-~[\phi ~(x~+~a)~+~\phi ~(x~-~a)]/2$
is equal to the sum of the even terms (even powers of $a$)
in the right-hand side of (5).
Solving the equations of motion by a recurrent procedure starting from the
local case, would lead to an expansion:
\equation
\phi ~(x)~~=~~\Sigma _{m=0}^{\infty}~\phi _m~(x)~a^{2m}
\endequation
which is valid for any nonlocal model, whether or not
it derives from a lattice. $\phi _0~(x)$ corresponds to the local, 
relativistic limit with exact Lorentz symmetry. 
If the model has solitons, in the relativistic limit 
their size will be $\gamma ^{-1} ~\Delta $ , 
where $\Delta $ is a characteristic
distance scale from soliton dynamics which cannot
be generated from just $a$ and the speed of light $c$~. Such a scale can 
be provided by a local potential $V~(\phi )$ which would need a
constant with space (or time) dimensionality in order to be compared with 
the second time derivative and with the space finite difference
(or with the second space derivative, in the continuum limit).
Then, a dimensionless parameter for power expansion would be: 
$\xi ~=~\alpha ~(a~\gamma _R)^2~\Delta ^{-2}$  
where $\alpha $ is basically the same coefficient
as before, accounting for the coefficients of the power expansion,
and $\gamma _R $ is the standard relativistic Lorentz factor. 
If the effective inverse squared Lorentz factor 
$\gamma ^{-2}$ is corrected by a power series of $\xi $ 
($\gamma ^{-2}~=~\gamma _R^{-2}~+~\gamma '~\xi ~+~...$, 
$\gamma '$ being a constant of order 1), we expect the 
departure
from standard relativity to play a leading role at 
energies above that for which $\gamma _R^{-2}~\approx ~
\alpha ~(a~\gamma _R)^2~\Delta ^{-2}$ , i.e. above $E~\approx
~m~c^2~\alpha ~^{1/4}~(a~\Delta ^{-1})^{-1/2}$ . Taking the values 
$\alpha ~\approx ~0.1$ , $m~\approx ~1~GeV/c^2$
and $\Delta ~\approx ~10^{-13}~cm$ , this 
energy scale corresponds to $E$ above $\approx ~
2.10^{19}~eV$ for $a~\approx ~10^{-33}~cm$~,
$2.10^{20}~eV$ for $a~\approx ~10^{-35}~cm$
and $\approx ~5.10^{17}~eV$ for
$a~\approx 10^{-30}~cm$ . For comparison, the maximum velocity 
in the previous deformed relativistic kinematics occurs 
at $E~\approx ~10^{19}~eV$ for $a~\approx ~10^{-33}~cm$ ,
$\approx ~10^{20}~eV$ for $a~\approx ~10^{-35}~cm$ and 
$\approx ~3.10^{17}~eV$ for $a~\approx 10^{-30}~cm$ . 
The departure from standard relativity, concerning the size 
and internal time scale of the soliton 
in the vacuum rest frame, indicates that its internal structure is
modified and we do not expect the usual parton model to hold any longer.

A rough, indicative illustration of the basic mechanism we just described
can be obtained by introducing a small perturbation in the $\Phi ^4$ soliton
theory. Starting from the equation:
\equation
c^{-2}~\partial ^2\psi /\partial t^2~~-~~\partial ^2\psi /\partial x^2~~~
=~~~2~\Delta ^{-2} ~\psi ~(1~-~\psi ^2)
\endequation
where $\Delta $ is the distance scale characterizing the soliton size
($x$ = space coordinate, $t$ = time coordinate), and 
writing down the one-soliton solution of this equation:
\equation
\psi (x~,~t)~~=~~\Phi (y)~~=~~tanh~(\lambda _0~y)
\endequation
where $y~=~x~-v~t$ , $v$ is the speed of the soliton, $\lambda _0~=
~\Delta ^{-1}~\gamma _R$ and, as before, $\gamma _R~=~(1~-~v^2~c^{-2})^{-1/2}$ ,
we can introduce a perturbation to the system by adding to the left-hand side of
(7) a term $-~(a^2/12)~\partial ^4\psi /\partial~x^4$ . The new differential
term in the equation corresponds to the term proportional to $a^4$ in the
expansion (5) , 
and will be compensated at the
first order in the perturbation by the replacement:
\equation
\Phi ~~\rightarrow ~~\Phi ~+~\epsilon ~\Phi~(1~-~\Phi ^2)
\endequation
where $\epsilon ~\propto ~a^2$ .
We furthermore replace $\lambda _0$ by a new coefficient $\lambda $ to
be determined form the perturbed equation. To first order in the perturbation,
we get the solutions:
\equation
\epsilon~~\simeq ~~1~-~\lambda ^2~\gamma _R^{-2}~\Delta ^2
\endequation
\equation
\lambda ^2~~\simeq ~~[3~\pm ~(1~-~4~a^2~\Delta ^{-2}~\gamma _R^4/3)^{1/2}]~
(1~+~a^2~\Delta ^{-2}~\gamma _R^4/6)^{-1}~\Delta ^{-2}~\gamma _R^2/4
\endequation
leading for $\epsilon ~\ll ~1$
to $(\Delta ~\lambda )^{-2}~\simeq ~\gamma _R^{-2}~+
~a^2~\Delta ^{-2}~\gamma _R^2/3$
in agreement with our previous discussion. Maximum values of $v$ 
and $\lambda $ would occur at 
$\gamma _R~=~2^{-1/2}~3^{1/4}~(\Delta ~a^{-1})^{1/2}$ ,
but our approximations do not apply at this value of $\gamma _R$ . Deformed 
kinematics can be obtained at the lowest order in the perturbation.
A simplified calculation, valid for a wide class of soliton models, 
could be as follows. At the first order in the perturbation with respect 
to special relativity, we start from an effective lagrangian 
for soliton kinematics: 
\equation
L~~=~~-~m~c^2~\gamma _R^{-1}~(1~-~\rho ~\gamma _R^4)
\endequation
where $\rho $ is a constant proportional to $a^2~\Delta ^{-2}$ , according to
(10) and (11). From this lagrangian, we derive the expression for the
generalized momentum: 
\equation
p~~=~~m~\gamma _R~v~(1~+~3~\rho ~\gamma _R^4)
\endequation
from which we can build the hamiltonian:
\equation
H~~=~~p~v~-~L~~=~~m~c^2~(\gamma _R~+~3~\rho ~v^2~c^{-2}~\gamma _R^5~-
~\rho ~\gamma _R^3)
\endequation
which 
leads to a deformed relativistic kinematics defined by the relation:
\equation
E~-~p~c~~=~~m~c^2~\gamma _R^{-1}~(1~+~v~c^{-1})^{-1}~-~\rho ~~m~c^2~
[3~v~c^{-1}~(1~+~v~c^{-1})^{-1}~+~1]~\gamma _R^3
\endequation
and, when expressed in terms of $p$ at $v~\simeq ~c$ , can be approximated by:
\equation
E ~-~p~c~~\simeq ~~m~c^2~(2~p)^{-1}~-~5~\rho ~p^3~(2~m^2~c)^{-1}
\endequation
where
the deformation term $5~\rho ~p^3~(2~m^2~c)^{-1}$
differs from that obtained from phonon mechanics
only by a constant factor $\eta ~\propto ~2~h^2~(2\pi ~m~c~\Delta)^{-2}$ . 
This factor can be close to 1
only if the mass of the extended system is directly related to its size 
through a quantum uncertainty relation. Such a situation corresponds to
reality for hadrons, but not for nuclei (or larger bodies) where both the mass
and the size increase simultaneously
as the system gets larger. In general, there seems to be no 
fundamental reason for $\eta $ to be exactly 1 , but we have no indication
that it should vary like the inverse square of the number of quarks forming
the hadron, as previously foreseen for composite models
(Gonzalez-Mestres, 1997c). 
The deformation of relativistic kinematics seems to be driven by soliton
dynamics and not by the superposition of deformed kinematics for free quarks.
From the above
heuristic example, both $\eta ~>~1$ and  $\eta ~<~1$ are possible but $\eta $
seems to naturally be $\approx ~1$ for hadrons and $\ll ~1$ for heavy
nuclei. Also, the value of $\alpha $ will a priori be different for different
hadrons.
A more precise estimate of the effect of non-locality in hadronic physics 
could be obtained by putting QCD on a lattice whose spacing would be the 
fundamental length $a$ but, due to the very large $\Delta~/a$
ratio, the lattice size would be far beyond numerical computation 
potentialities. We shall not attempt here to introduce nonlocality in other
theoretical approaches to the structure of hadrons. Looking at the low-speed 
limit of (15), we find a renormalization of the critical speed 
parameter $c$ ,
$\delta c$ , such that $\delta c~c^{-1}~
\approx ~(\Delta /a)^{-2}~\approx ~10^{-40}$ for hadrons if
$a~\approx ~10^{-33}~cm$ and $\Delta ~\approx ~10^{-13}~cm$ . This effect is 
much smaller than the effects contemplated by other authors (e.g. Coleman and
Glashow, 1997) and cannot be excluded by existing data.
 
Above the critical momentum scale (in the vacuum rest frame)
defined by the previous arguments,
we can imagine two scenarios for very high-energy hadrons:

- a) As the deformation term $\alpha ~(k~a)^3/2$ in (3) becomes much larger
than the mass term $(2\pi ~a)^2~h^{-2}~m^2~c^2$ , 
the internal structure of the composite system evolves and
the soliton may no longer
be stable. For hadrons, this would lead to a scenario with free quarks 
and gluons at
high enough energies in vacuum rest frame
(kinematic deconfinement). Quarks and gluons with energies just above the 
threshold for kinematic deconfinement would not be able
to transfer energy to slower 
particles, but could form with them fast aggregates. 

- b) Confinement persists, and the departure from standard relativity
becomes stronger as momentum increases. 
As $k$ becomes $\approx ~a^{-1}$ ,
the soliton may slower down and possibly recover a size similar to its
low-momentum rest size. The dynamics well above the critical momentum scale 
(the scale where the deformation term becomes of the same order as the mass 
term) is
likely to be model-dependent, and no general conclusion can {\it a priori}
be drawn in the lack of a precise dynamical description of hadrons.
The assumption that has the same value
for leptons, gauge bosons and hadrons is possibly a qualitatively reasonable
hypothesis, but not the exact solution of the problem.
 
In the case of nuclei, where nucleons are not confined but just bound by 
strong interactions,  
the question of stability and
internal structure of nuclei at very high energy is indeed a crucial one.
It can be checked that relative corrections to the boosted transverse energy
are proportional to $\approx ~\alpha ~(k~a)^2$ and remain very small as long
as $k~a~\ll ~1$ . If, inside the nuclei, the nucleons must be considered as
individual particles, the value of $\alpha $ for a given nucleus must be
corrected by a factor $N^{-2}$ , where $N$ is now the number of
nucleons. If the above description based on soliton models applies to
nuclei, and taking $\Delta $ to be the same as for single nucleons (the 
nucleus would be a multi-soliton bound state), the correction factor
would be proportional to $m^{-2}$ ($m$ = mass of the nucleus). With $a~\approx
~10^{-33}~cm$ , and assuming spontaneous disintegration of the nucleus
at rest to
be protected by a binding energy $\approx ~10^{-2}~m~c^2$ , this version of 
nuclear deformed
relativistic kinematics would still be able to inhibit photodisintegration 
by cosmic background photons
at energies above $\approx ~N^{2/3}~10^{19}~eV$ if $\alpha ~\approx ~0.1$ 
for protons. 
However, if $\alpha ~=~\alpha _{el}$ for elementary particles and $\alpha ~
\approx ~\alpha _{el}~N^{-2}$ for a nucleus made of $N$ nucleons, a maximum 
energy attainable by the nucleus is $E~\approx ~2~a^{-1/2}~m^{1/2}~
\alpha _{el}^{-1/4}~c^{3/2}~h^{1/2}~(2~\pi )^{-1/2}$ , i.e. $E~\approx ~
N^{1/2}~10^{19}~eV$ for $\alpha _{el}~\approx ~0.1$ and 
$a~\approx ~10^{-33}~cm$ . Above these energies, the nucleus can spontaneously
radiate very high-energy photons. With the same values 
for $\alpha _{el}$ and $a$ , synchrotron radiation
would be inhibited only above 
$E~\approx ~a^{-1/2}~m^{1/2}~\alpha _{el}^{-1/4}~
c^{3/2}~h^{1/2}~(2~\pi )^{-1/2}$ , i.e. above $E~\approx ~0.5~N^{1/2}~
10^{19}~eV$ . The physics outcome for nuclei 
in the GZK cutoff region would therefore be sensitive to the details of
nuclear parameters, source distances to earth and acceleration mechanisms.
Contrary to 
our previous estimates based on
a universal value of $\alpha $ for elementary particles and nuclei
(Gonzalez-Mestres, 1997a and 1997b), unstable nuclei would not necessarily 
become stable at very high energy; possible departures from standard 
relativistic lifetimes (if any) will occur only at higher energies, and will
be very sensitive to the details of deformed relativistic kinematics.

Another possibility
would be to consider the nucleus as a single particle 
with a single collective 
motion to which deformed relativistic kinematics applies
with the same value of $\alpha $ as for leptons, gauge bosons and hadrons.
Indeed, for the nucleus to be stable, all nucleons should propagate at 
exactly the
same speed and therefore cannot be considered 
at very high energy as individual constituents carrying
variable fractions of energy and momentum
(even within the limits set by the strength of the interaction, the
potential and kinetic energies of nucleons inside a nucleus at rest
being in any case much smaller than the nucleon rest energies).  
Again, only the overall system undergoes
an independent, fast motion.
Then, the calculations presented
in previous papers (e.g. Gonzalez-Mestres, 1997a and 1997b) would apply without
further modifications. 
In particular, fission and decays changing the nature of the nucleus 
($\alpha $ , $\beta $ ...) would be forbidden above well-defined critical 
energies. A natural prediction of this scenario would be the existence,
at very high momentum, of
very heavy nuclei impossible to produce in the laboratory. 
Furthermore, besides the expected
fall of elastic, multiparticle and total
cross sections (Gonzalez-Mestres, 1997d), the new kinematics may give rise
to fundamentally new kinds of reactions.
If the value of $\alpha $ is universal for all particles and nuclei,
a very high-energy nucleon or nucleus can fusion with a low-energy nuclear
target producing a heavier nucleus which would be stable at this value of
momentum. At high enough energy, this reaction, which 
would release an amount of energy
of the order of the  
energy of the target (including its rest energy), 
can happen several times, leading eventually to a very heavy
stable nucleus-like system not described in
standard nuclear tables. 
Contrary to quark nuggets, these objects would in general
be unstable at low energy and cannot be found at nonrelativistic
speeds. Experiment seems thus, in principle,
to be able to discriminate between different scenarios
for very high-energy nuclear physics, in particular concerning the crucial
question of the values of $\alpha $ for hadrons and nuclei as compared to
leptons and gauge bosons.

To try to settle the discussion, it seems relevant 
to consider extrapolations to atoms, molecules
and macroscopic systems. It is not obvious that such extrapolations exist,
and that such objects can be accelerated with respect to the vacuum rest frame 
(without being disintegrated) to speeds close to the above 
mentioned maximum, or that they can reach the kinematical region where 
the deformation of relativistic kinematics dominates over the mass
term. If this is
the case, there is no compelling argument to apply to these systems the
same deformed kinematics as to "elementary" particles, hadrons or nuclei.
The size of and atom being much larger than that of a nucleus, and the binding
energy much lower, it may seem reasonable to consider the atomic 
nucleus and the 
electrons as separate particles to which deformed relativistic kinematics
applies individually. This seems also to follow from consistency requirements:
a body with mass $10^{25}~eV~c^{-2}$ ($\approx 10^{-8}~gm$)
moving at galactic speed would most 
likely have with respect to the vacuum rest frame (e.g. assuming it to be
close to that defined by cosmic microwave background radiation) a momentum
$\approx 10^{22}~eV~c^{-1}$ . It would obviously be erroneous to apply 
single-particle deformed relativistic kinematics to such an object
with the same values of $\alpha $ as for elementary particles. 
At this stage, two different simplified approaches can be considered:

- {\it Model i)} . Due to the very large size of atoms, as
compared to nuclei, the transition from 
nuclear to atomic scale appears as a reasonable point to stop considering
systems as "elementary" from the point of view of deformed relativity.
$\alpha $ would then have a universal value for nuclei and simpler objects,
but not for atoms and larger bodies. 

- {\it Model ii)} . The example with $\Phi ^4$ solitons suggests that 
hadrons can have values of $\alpha $ close to that of leptons and 
gauge bosons, and the
transition may happen continuously at fermi scale,
when going from nucleons to nuclei. Then, the value of $\alpha $ would be
universal (or close to it) for leptons, gauge bosons and hadrons but follow
a $m^{-2}$ law for nuclei and heavier systems,
the nucleon mass
setting the scale.  

Experimental tests should be performed and equivalent dynamical systems 
should be studied.
However,  {\it Model i)} would lack a well-defined criterium to separate
systems to which the deformed relativity applies with the same value of 
$\alpha $ as for leptons and gauge bosons from those to which this kinematics
cannot be applied, and to characterize the transition between the two regimes. 
The above obtained $m^{-2}~\Delta ^{-2}$ dependence of
the coefficient of the deformation term for extended objects,
as described in {\it Model ii)}, seems to provide a continuous transition
from nucleons to heavier systems, naturally
filling this gap. On the other hand, a closer analysis reveals that there is
indeed a discontinuity between nuclei and atoms, as foreseen in 
{\it Model i)} . As long as the deformation term 
in electron kinematics can be neglected
as compared to the electron mass term, we can consider that most of the
momentum of an atom is carried by the nucleus and {\it Model ii)} may provide
a reasonable description of reality. But, when the electron mass term becomes
small as compared to the part of the energy it would carry in a parton
model of the atom, such a description becomes misleading. To have the same
speed as a nucleon, the electron must then carry nearly the same energy and 
momentum.  
We therefore propose a modified version of {\it Model ii)} 
with $\Delta \approx~10^{-13}~cm$ from hadrons and nuclei where, for atoms 
and larger neutral systems, the coefficient of the deformation term would be
corrected by a factor close to 1 at low momentum and 
to 4/9 at high momentum
if the number of neutrons is equal to that of protons. This model is
obviously approximate and should be completed by a detailed dynamical
calculation that we shall not attempt here. It assumes that electrically
neutral bodies can reach very high energies per unit mass, which is
not obvious: spontaneous ionization may occur at speeds (in the
vacuum rest frame) for which 
the deformation term in electron kinematics becomes larger 
than its mass term.

Then, for bodies heavier than hadrons, the effective value of $\alpha $ would 
decrease essentially like $m^{-2}$ . Applying a similar mass-dependence
to the $\kappa $ parameter of a different 
deformed Poincar\'e algebra considered by
previous authors (Bacry, 1993 and references therein), i.e. $\kappa~ \propto ~
m$ for large bodies, yields the relation:
\equation
F~(M_0~,~E_0)~~=~~F~(M_1~,~E_1)~~+~~F~(M_2~,~E_2)
\endequation
with:
\equation
F~(m~,~E )~~=~~2~\kappa ~(m)~~sinh~[2^{-1}~\kappa ^{-1}~(m)~E ]
\endequation
where $M~=~M_1~+~M_2$ , $\kappa (m)$ is our above mass-dependent version  
of the
$\kappa $ parameter of the deformed Poincar\'e algebra used by these authors,
and $E_0$ is the energy of a system with mass $M$ made of two non-interacting 
subsystems of energies $E_1$ and $E_2$ and with masses $M_1$ and $M_2$ . 
Defining mass as an additive parameter, the rest energy $E_{i,rest}$ 
(in the vacuum rest frame) of particle $i$ ($i~=~0~,~1~,~2$)
is given by the equation:
\equation
M_i~c^2~~=~~2~\kappa ~(M_i)~~sinh~[2^{-1}~\kappa ^{-1}~(M_i)~E_{i,rest}]
\endequation
and tends to $M_i~c^2$ as $~\kappa ~(M_i)~\rightarrow ~\infty$ .
Equations (17) and (18) lead to additive relations for the energy of macroscopic
objects if the proportionality rule $\kappa ~(m)~\propto ~m$ 
is applied. From our 
previous discussion with a different deformation scheme, such a choice
seems to naturally agree with physical reality. Then, 
contrary to previous claims (Bacry, 1993; Fernandez, 1996),
the rest energies of  
large systems would be additive and no macroscopic effect on the total mass
of the Universe would be expected.


\section{Experimental considerations}

Definitely, the main and most fundamental
physics outcome of very high-energy cosmic-ray 
experiments involving particles and nuclei
may eventually be the test of special relativity. If Lorentz
symmetry is violated at Planck scale, the highest-energy cosmic ray events
may, if analyzed closely and with the expected high statistics from future
experiments, provide a detailed check of different models of deformed
relativistic kinematics. If some of the highest-energy cosmic rays are nuclei,
their study would yield crucial information on Planck-scale corrections to the 
kinematics of composite objects. If Lorentz symmetry turns out to be violated,
the comparison between the properties of
nucleons and those of different nuclei would be crucial to understand how
deformed Poincar\'e algebras possibly 
apply to matter. Indeed, a detailed fit to data would involve and check
all the relevant parameters. Although the models we consider 
predict the new phenomena to occur in all cases at energies above $\approx ~
10^{17}~eV$ for hadrons and nuclei, extrapolation over several orders of 
magnitude is involved and many theoretical uncertainties remain. 
Furthermore, until now we have been interested only in observable effects
at leading level, but small deviations from standard relativistic 
values of physical parameters may also be detectable (although, in the 
models we considered, the energy dependence of the predicted new effects
is always very sharp since a fourth power of energy is often involved in
the calculations).
Therefore,
it would be most crucial to find an efficient way 
to test special relativity for cosmic rays in
the $10^{16}~-~10^{17}~eV$ range with the help of LHC
using the potentialities of the FELIX experiment (e.g. Eggert, Jones and 
Taylor, 1997). Such a test is actually compulsory before trying to consistently
compare LHC and cosmic-ray data for other practical purposes.

Lorentz symmetry violation
prevents naive extrapolations from reactions between two particles 
with equal, opposite momenta in the vacuum rest frame 
(similar to colliders) to reactions where 
the target is at rest in this frame
(similar to cosmic-ray events). {\it Assuming the earth to move slowly with
respect to the vacuum rest frame} (for instance, if the "absolute" frame
is close to that defined by the requirement of cosmic microwave background
isotropy),
the described kinematics leads, at the
highest observed cosmic ray energies, to a departure from the parton model
that can strongly influence cascade development at very high
energy. 
Other effects, such as a longer $\pi ^0$ lifetime as compared to standard
relativity or the possibility that the primary hadron be a neutron,
would increase the interest and originality of very high-energy
showers.
Furthermore, we 
predict (Gonzalez-Mestres, 1997d)
the existence of 
a maximum energy deposition for high-energy
cosmic rays in the atmosphere, in the rock or in a given underground or
underwater detector. Well below Planck energy, a very high-energy
cosmic ray would not necessarily
deposit most of its energy in the atmosphere: its energy 
deposition decreases for energies above a transition 
scale, far below 
the energy scale associated to the fundamental length.
The maximum allowed
momentum transfer in a single collision occurs at an energy just below
$E_{lim}~\approx ~(2\pi~)^{-2/3}~\alpha ^{-1/3}~(E_T~a^{-2}~h^2~c^2)^{1/3}$ , 
where $E_T$ is the energy of the
target in the vacuum rest frame.
For $E$ above $E_{lim}$ 
($\approx 4.10^{22}~eV$ if the target is a non-relativistic nitrogen or oxigen
nucleus, $\alpha ~\approx ~0.1$ for the 
cosmic ray and $a~\approx ~10^{-33}~cm$), 
the allowed longitudinal momentum transfer falls, typically, 
like $p^{-2}$ 
(obtained differentiating
the term $\alpha ~k^2~a^2~p~c/2$). 
At energies around 
$E_{lim}$ , 
the cosmic ray will in our scenario undergo several 
scatterings in the atmosphere and still lose there most of its energy,
possibly
leading to unconventional longitudinal cascade development profiles
that could be studied in detail by very large-surface air shower detectors
like the AUGER observatory (AUGER Collaboration, 1997).
Above $E_{lim}$~, the cosmic 
ray can indeed cross the
atmosphere keeping most of its momentum and energy and deposit its energy 
in the rock or in water, or possibly reach and underground or underwater
detector. Thus, some unconventional
cosmic-ray events of apparent energy far below 
$10^{20}~eV$ , 
as seen by earth-surface
(e.g. air shower), underground or underwater detectors,
may actually be originated by extremely high-energy cosmic rays well above
this energy scale but interacting weakly because of deformed kinematics. 

Nuclei would play a crucial role in the analysis of these phenomena, due to
their sensitivity (through the values of the $\alpha $ parameter) to the basic
mechanisms of Lorentz symmetry violation and (hopefully) to the possibility to
explore different nuclear masses in a high-statistics experiment devoted to the
highest-energy cosmic rays. 

Constraints on the fundamental length $a$ were derived
(Gonzalez-Mestres, 
1997d) assuming deformed relativistic kinematics with universal values of
$\alpha $ and $c$ , and full-strength Lorentz symmetry violation at the
fundamental length scale. In spite of the criticism presented in this paper, 
such estimates remain useful. 
The
combined absence of GZK cutoff and existence of $\approx 10^{20}~eV$ energy
deposition from cosmic rays in the atmosphere 
lead to $a$ in the range $10^{-35}~cm~<~a~<~10^{-30}~cm$ 
(energy scale between $10^{16}$ and $10^{21}$ $GeV$). The lower bound comes
from the requirement that the violation of local Lorentz invariance at the
fundamental length scale be able to influence particle interactions
at the $10^{19}~-~10^{20}~eV$ energy scale strongly
enough to suppress the GZK cutoff. The
upper bound is derived from the existence of events with
$\approx 10^{20}~eV$ energy deposition in the atmosphere (Linsley, 1963;
Lawrence, Reid and Watson, 1991; Afanasiev et al., 1995; 
Bird et al., 1994; Yoshida et al., 1995; Hayashida et al., 1997).
Then, the departure from the
parton model is expected to occur at an energy scale between
$\approx ~3.10^{17}~eV$ (for $a~\approx ~10^{-30}~cm$) and $\approx 10^{20}~eV$ 
(for $a~\approx ~10^{-35}~cm$), in all cases at energies below the highest
measured cosmic ray energies. Assuming a universal value of 
$\alpha ~\approx 0.1$ for
leptons, gauge bosons and hadrons, and the highest-energy cosmic rays to be 
nucleons, it would also be possible to fit the data with a model where 
the GZK phenomenon 
acts at energies below $10^{20}~eV$ but is inhibited above this energy.
The best value of $a$ for this purpose would be $\approx 3.10^{-36}~ cm$ ,
slightly below the range considered in our previous estimates.  
It would lead to a failure of the parton model
at energies above $\approx 2.10^{20}~eV$ , and to longer $\pi ^0$ lifetimes
(inhibiting electromagnetic showers and favouring strong interactions leading
to pairs of muons)
above $\approx 2.10^{19}~eV$ (this phenomenon occurs above $E~\approx ~
10^{18}~eV$ if $a~\approx ~10^{-33}~cm$). Energy losses in acceleration 
processes would be inhibited only at higher energies than previously 
considered, but the phenomenon would still be crucial for the highest-energy
cosmic rays.
If the hypothesis of full-strength Lorentz symmetry violation at the
fundamental length scale is abandoned, the roles of $\alpha $ and $a$
are expressed by a single parameter, $\alpha ~a^2$ : then, $a~\approx ~
3.10^{-36}~cm$ and $\alpha ~\approx ~0.1$ would be equivalent to
$a~\approx ~10^{-33}~cm$ and  $\alpha ~\approx ~10^{-6}$ .
However, since $\alpha $ is not larger than $\approx ~0.1$ , $a$ cannot be
smaller than $\approx ~3.10^{-36}~cm$ . If Lorentz symmetry violation
at the fundamental length scale is very weak, this scale may occur well 
below Planck scale without contradicting data: $a~\approx ~10^{-33}~cm$ 
and $\alpha ~\approx ~0.1$ would be equivalent to $a~\approx ~10^{-26}~cm$
(just below length scale associated to the highest cosmic ray energies)
and $\alpha ~\approx ~10^{-15}$ . Only the detection of cosmic rays with
energies above
$E~\approx ~10^{21}~eV$ would possibly allow to exclude a fundamental
length scale $a~\approx ~10^{-26}~cm$ .
Apart from this kind of ambiguity, a detailed analysis of the structure of 
cosmic-ray events at energies above $\approx ~10^{17}~eV$ would allow to 
discriminate between different scenarios. 
Future colliders with energies above LHC energies
would allow for direct comparison (check of Lorentz invariance)
between collider events and cosmic-ray events in this energy range. 

Then, very high-energy accelerator and cosmic-ray experiments would indeed
be complementary research lines: the results
of both kinds of experiments would not be equivalent up to Lorentz 
tranformations.
A $p~-~p$ collider at $\approx 700~TeV$ per beam, accelerating also nuclei,
could make possible direct tests of
Lorentz symmetry violation, comparing collisions at the accelerator with 
collisions between a $\approx 10^{21}~eV$ proton 
of cosmic origin and a proton or nucleus from the
atmosphere (an important step in this direction could be taken by the VLHC
machine, Very Large Hadron Collider reaching $\approx 100~TeV$ per beam,
which corresponds to cosmic proton energies $\approx 2.10^{19}~eV$) . 
Simultaneously, other kinds of tests may be possible through the
lifetimes and decay products of very high-energy unstable particles  
(Gonzalez-Mestres, 1997a and 1997b) in the cosmic-ray events 
producing the highest-energy secondaries (e.g. by looking in some way at
the first $\pi ^0$~'s 
produced). 
We would be confronted to a new situation, 
contrary to conventional expectations, 
if the cosmic rays at the highest possible
energies interact more and more weakly with matter because of kinematical
constraints. The existence of a maximum energy of events generated in
the atmosphere would not correspond to a maximum energy of 
incoming cosmic rays.
Unconventional events   
originated by such particles 
may have been erroneously 
interpreted as being associated to cosmic rays of much lower energy.
New analysis seem necessary, as well as 
new experimental designs 
using perhaps in coincidence very large-surface 
detectors devoted to 
interactions in the atmosphere (like AUGER) with 
very large-volume underground or underwater detectors (like AMANDA)
and with balloon or
satellite experiments able to closely analyze the early cascade development.
\vskip 6mm
{\bf References}
\vskip 4mm
\noindent
Afanasiev, B.N. et al., Proc. of the $24^{th}$ International Cosmic Ray 
Conference, Rome, Italy, Vol. 2 , p. 756 (1995).\newline
Anchordoqui, L., Dova, M.T., G\'omez Dumm, D. and Lacentre, P.,
{\it Zeitschrift f\"{u}r Physik C} 73 , 465 (1997).\newline
AUGER Collaboration, "The Pierre Auger Observatory Design Report"
(1997).\newline
Bacry, H., Marseille preprint CPT-93/P.2911 , available from KEK 
database.\newline
Bird, D.J. et al., {\it Phys. Rev. Lett.} 71 , 3401 (1993).\newline
Bird, D.J. et al., {\it Ap. J.} 424 , 491 (1994).\newline 
Blokhintsev, D.I. and Kolerov, G.I., {\it Nuovo Cimento} 
34 , 163 (1964).\newline
Blokhintsev, D.I., {\it Sov. Phys. Usp.} 9 , 405 (1966).\newline
Coleman, S. and Glashow, S.L. , {\it Phys. Lett. B} 405 , 249 (1997).\newline
Dova, M.T., Epele, L.N. and Hojvat, C., Proceedings of the 25$^{th}$ 
International Cosmic Ray Conference (ICRC97), Durban, South Africa, Vol. 7 ,
p. 381 (1997).\newline
Eggert, K., Jones, L.W. and Taylor, C.C., Proceedings of ICRC97 , Vol. 6 , 
p. 25 (1997).\newline
Fernandez, J.,  {\it Phys. Lett. B} 368 , 53 (1996).\newline 
Gonzalez-Mestres, L., "Physical and Cosmological Implications of a Possible
Class of Particles Able to Travel Faster than Light", contribution to the
28$^{th}$ International Conference on High-Energy Physics, Warsaw July 1996 .
Paper hep-ph/9610474 of LANL (Los Alamos) electronic archive (1996).\newline
Gonzalez-Mestres, L., "Vacuum Structure, Lorentz Symmetry and Superluminal
Particles", paper physics/9704017 of LANL 
electronic archive (1997a).\newline
Gonzalez-Mestres, L., "Absence of Greisen-Zatsepin-Kuzmin Cutoff 
and Stability of Unstable Particles at Very High Energy,
as a Consequence of Lorentz Symmetry Violation", Proceedings of ICRC97 ,
Vol. 6 , p. 113 . Paper 
physics/9705031 of LANL electronic archive (1997b).\newline
Gonzalez-Mestres, L., "Possible Effects of Lorentz Symmetry Violation on the
Interaction Properties of Very High-Energy Cosmic Rays", paper
physics/9706032 of LANL electronic archive (1997c).\newline
Gonzalez-Mestres, L., "Lorentz Symmetry Violation and Very High-Energy
Cross Sections", paper physics/9706022 of LANL electronic archive (1997d).
\newline
Gonzalez-Mestres, L., "Space, Time and Superluminal Particles",
paper physics/9702026 of LANL electronic archive (1997e).\newline
Gonzalez-Mestres, L., "Lorentz Invariance and Superluminal Particles",
paper physics/9703020 of LANL electronic archive (1997f).\newline
Greisen, K., {\it Phys. Rev. Lett.} 16 , 748 (1966).\newline
Hayashida, N. et al., {\it Phys. Rev. Lett.} 73 , 3491 (1994).\newline
Hayashida, N. et al., Proceedings of ICRC97 , Vol. 4 , p. 145 (1997).\newline
Lawrence, M.A., Reid, R.J.O. and Watson, A.A., {\it J. Phys. G }, 17 , 773
(1991).\newline
Linsley, J., {\it Phys. Rev. Lett.} 10 , 146 (1963).\newline
R\'edei, L.B., {\it Phys. Rev.} 162 , 1299 (1967).\newline
Yoshida, S. et al., Proc. of the $24^{th}$ International Cosmic Ray 
Conference, Rome, Italy, Vol. 1 , p. 793 (1995).\newline
Zatsepin, G.T. and
Kuzmin, V.A., {\it Pisma Zh. Eksp. Teor. Fiz.} 4 , 114 (1966).
\end{document}